# A Non-Relativistic Argument Against Continuous Trajectories of Particles

*by Sofia D. Wechsler*

**Abstract**
A thought-experiment is described and the probability of a particular type of results is predicted according to the quantum formalism. Then, the assumption is made that there exists a particle that travels from the source to one of the detectors, along a continuous trajectory. A contradiction appears: for agreeing with the quantum prediction, the particle has to land at once on two space-separated detectors. Therefore, the trajectory of the particle – if it exists – cannot be continuous.

**Abbreviations**
3-wave = a product of states of three particles
dB-B = de Broglie-Bohm
$D_1 \& D_2 \& D_3$ = joint detection, one in the detector $D_1$, one in $D_2$, and one in $D_3$
EBS = end-beam-splitter
QM = quantum mechanics
w-f = wave-function
w-p = wave-packet

## 1. Introduction

The opinion predominating in the quantum community is that the quantum system has a double nature, namely, in which-way experiments it behaves as a particle, and in interference experiments it behaves as a wave. However, even in interference tests, if in the interference region is placed a sensitive plate, in each trial of the experiment a dot appears on the plate. Therefore the picture that seems to prevail is that of a particle carried by a wave-packet (w-p), and it's the particle that impresses the plate.

However, if one rejects the possibility that the particle may disappear from the space into nothing, or appear from nothing, one should admit that the particle follows a continuous trajectory from the region where the wave-function (w-f) is prepared, to the detector.

The formalism of the de Broglie-Bohm (dB-B) mechanics [1], [2], [3], [4] is built on this idea. It supplements the quantum mechanics (QM) formalism with a sub-quantum structure consisting of a particle floating inside the w-f. The particle is supposed to follow a *continuous* trajectory, along which the position and linear momentum are simultaneously well defined. Though, according to the QM, simultaneous well defined position and linear momentum is in contradiction with the uncertainty principle. The dB-B mechanics attributes the uncertainty principle to lack of knowledge, i.e. a measurement of position perturbs the linear momentum and vice-versa.

Many physicists embraced the dB-B interpretation of QM because they saw in it a way to get rid of the enigmatic principle of collapse. Though, about seventy years from de Broglie's Ph.D. thesis and theory of waves [1, 2], and about forty years after the publication of Bohm's famous articles [3], [4], L. Hardy found that the Bohmian trajectories are not relativistically covariant [5]. It cast the first strong doubt on the dB-B mechanics as explained in detail by Berndl et al. [6] (see also [7] for a simpler explanation). The supporters of the dB-B mechanics had to accept that its formalism is valid only in the non-relativistic domain.

The problem found in [5], [6], [7] doesn't stem from the dB-B formulas, contrary to the opinion of the authors of the ESSW thought experiment [8], [9][1]; it stems from the basic assumption of continuous trajectories. This is the central idea of the proof in the present work, that quantum objects don't follow continuous paths.

Ruling out continuous trajectories doesn't yet rule out the possibility that substructure particles exist. S. Gao [10], [11], [12], came with the idea that since different parts of the w-f of an electron do not repel one another, there can exist a particle that jumps from one w-p to another, and the time of sojourning in a w-p is proportional to the absolute square of the amplitude of the w-p. Unfortunately, his idea is accompanied by many serious problems, as discussed in [13].

A totally opposite interpretation of the QM was proposed by Ghirardi, Rimini and Weber [14], and later on joined Pearle [15], [16], and Gisin [17]. This interpretation denies completely the particle concept, and takes the collapse of the w-f as a real fact. The main assumption in [14], [15], [16], [17], is that the collapse occurs when the studied quantum system interacts with so many particles that the total conglomerate of perturbed particles represents a macroscopic object. From the famous example with Schrödinger's cat, is known, and quite generally accepted, that there is no such thing as a quantum superposition of macroscopic states of an object. Though, neither this interpretation of QM lacks problems, as shown in [18].

The present work aims exclusively at ruling out the possibility of continuous trajectory for substructure particles. If there weren't problems with Gao's idea, this work would have represented an argument in support of that idea.

In continuation, section 2 describes the thought-experiment and section 3 derives the quantum predictions for the statistics of a particular type of results. Section 4 introduces the concept of particle traveling with a w-p and the requirement of continuous trajectories, pointing to a contradiction. Section 5 contains a discussion and the conclusion.

## 2. The experiment

The thought-experiment presented below is a variant of an experiment of Tan, Walls, and Collett, [19], and bears resemblance with an application of [19] by L. Hardy, [20], which in fact inspired the present work.

Consider a one particle wave-function emitted from a source S – figure 1. The w-p passes through a beam-splitter BS, which reflects 1/3 from the intensity of the incident beam. The transmitted w-p encounters a second beam-splitter, BS', which reflects half of the intensity of the incident beam and transmits the other half. In continuation, each w-p passes through a phase-shifter by $\theta_1$, $\theta_2$, and $\theta_3$, on the path **a**, **b**, and **c**, respectively, then meets an end-beam-splitter (EBS) identical to BS'. The three paths from BS to the EBSs are of equal optical lengths, the only differences in the phases with which the w-ps meet the EBSs, being caused by the phase-shifts. The resulting wave-function is therefore

---

[1] In their work, Englert, Scully, Süssmann, and Walther, calculated the trajectories of an atom according to Bohm's mechanics and found a contradiction with the w-f predicted by the QM.

$$|\psi\rangle = -\frac{1}{\sqrt{3}}\left(e^{i\theta_1}|1;\mathbf{a}\rangle|0;\mathbf{b}\rangle|0;\mathbf{c}\rangle + |0;\mathbf{a}\rangle e^{i\theta_2}|1;\mathbf{b}\rangle|0;\mathbf{c}\rangle + |0;\mathbf{a}\rangle|0;\mathbf{b}\rangle e^{i\theta_3}|1;\mathbf{c}\rangle\right), \tag{1}$$

On the other side of each EBS impinges a local oscillator, a coherent beam of particles of the same type as those from the source S,

$$|\alpha_j\rangle = \mathcal{N}\left(|0\rangle + q|1;\mathbf{e}_j\rangle + qp/2;\mathbf{e}_j\rangle\ldots\right), \quad j = 1, 2, 3, \tag{2}$$

where $\mathcal{N}$ is the normalization factor, $q$ and $p$ are complex numbers with $|q|<1$, $|p|<1$, and $\mathcal{N}$, $q$, $p$, are the same for all three coherent beams. The total wave-function at this step is

$$|\Psi\rangle = \frac{\mathcal{N}^3}{\sqrt{3}}\left(e^{i\theta_1}|1;\mathbf{a}\rangle|0;\mathbf{b}\rangle|0;\mathbf{c}\rangle + |0;\mathbf{a}\rangle e^{i\theta_2}|1;\mathbf{b}\rangle|0;\mathbf{c}\rangle + |0;\mathbf{a}\rangle|0;\mathbf{b}\rangle e^{i\theta_3}|1;\mathbf{c}\rangle\right)\prod_{j=1}^{3}\left(|0\rangle + q|1;\mathbf{e}_j\rangle + qp/2;\mathbf{e}_j\rangle\ldots\right). \tag{3}$$

*Conventions and definitions*:

1) Since the expressions written according to the 2nd quantization are very long, we will omit from now on in the formulas the w-ps of vacuum, $|0;\cdot\rangle$;

2) for shortening the discourse, a product of three w-ps will be named "3-wave";

3) we will name the particle exiting the source S, "S-particle";

4) a joint detection by $D_1$, $D_2$, and $D_3$, we will be denoted as $D_1\&D_2\&D_3$.

We will be interested only in those trials of the experiment which end with the result $D_1\&D_2\&D_3$.

Opening the parentheses in (3) and writing explicitly only the relevant terms, one gets, [21],

$$|\Psi\rangle = \frac{\mathcal{M}}{\sqrt{3}}\left\{\left(e^{i\theta_1}|1;\mathbf{a}\rangle|1;\mathbf{e}_2\rangle|1;\mathbf{e}_3\rangle + e^{i\theta_2}|1;\mathbf{e}_1\rangle|1;\mathbf{b}\rangle|1;\mathbf{e}_3\rangle + e^{i\theta_3}|1;\mathbf{e}_1\rangle|1;\mathbf{e}_2\rangle|1;\mathbf{c}\rangle\right) + \ldots\right\}, \tag{4}$$

$$\mathcal{M} = \mathcal{N}^3 q^2, \tag{5}$$

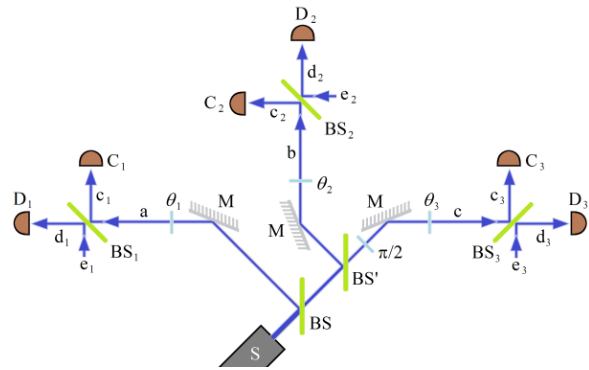

Figure 1. Schematic configuration of an experiment with three-particle entanglement.
See explanations in the text.

At the EBSs take place the transformations

$$|1;\mathbf{u}_j\rangle \to (1/\sqrt{2})(i|1;\mathbf{c}_j\rangle + |1;\mathbf{d}_j\rangle), \quad |1;\mathbf{e}_j\rangle \to (1/\sqrt{2})(|1;\mathbf{c}_j\rangle + i|1;\mathbf{d}_j\rangle), \quad j=1,2,3. \tag{6}$$

where $\mathbf{u}_1 = \mathbf{a}$, $\mathbf{u}_2 = \mathbf{b}$, and $\mathbf{u}_3 = \mathbf{c}$. Introducing (6) in (4) the wave function takes the form

$$|\Theta\rangle = \frac{\mathcal{M}}{\sqrt{24}}[(\sum_{m=1}^{3}e^{i\theta_1})(-i\prod_{m=1}^{3}|1;\mathbf{c}_m\rangle + \prod_{m=1}^{3}|1;\mathbf{d}_m\rangle) \\ + \sum_{\substack{j\neq k\neq l\neq j \\ \in\{1,2,3\}}}(e^{i\theta_j} + e^{i\theta_k} - e^{i\theta_l})(|1;\mathbf{c}_j\rangle|1;\mathbf{c}_k\rangle|1;\mathbf{d}_l\rangle - i|1;\mathbf{d}_j\rangle|1;\mathbf{d}_k\rangle|1;\mathbf{c}_l\rangle)] + \ldots \tag{7}$$

## 3. Quantum statistics of triple detections

The probabilities of joint clicks of the form one click beyond each EBS, are:

$$\text{Prob}(D_1 \& D_2 \& D_3) = \text{Prob}(C_1 \& C_2 \& C_3) = \frac{|\mathcal{M}|^2}{24}\left|e^{i\theta_1} + e^{i\theta_2} + e^{i\theta_3}\right|^2, \tag{8}$$

$$\text{Prob}(C_j \& C_k \& D_l) = \text{Prob}(D_j \& D_k \& C_l) = \frac{|\mathcal{M}|^2}{24}\left|e^{i\theta_j} + e^{i\theta_k} - e^{i\theta_l}\right|^2, \tag{9}$$

where $j,k,l \in \{1,2,3\}$, $j \neq k \neq l \neq j$.

One can infer from (8) and (9) the following:

1) The fact that the formulas (8) and (9) emerge from (4) implies that *three-particle-interference* was obtained. The RHS of (4) was written *for each trial*, and means that *in each trial* of the experiment all the three 3-waves $|1;\mathbf{a}\rangle|1;\mathbf{e}_2\rangle|1;\mathbf{e}_3\rangle$, $|1;\mathbf{e}_1\rangle|1;\mathbf{b}\rangle|1;\mathbf{e}_3\rangle$, and $|1;\mathbf{e}_1\rangle|1;\mathbf{e}_2\rangle|1;\mathbf{c}\rangle$ are present. In consequence, in each trial are present the w-p $|1;\mathbf{a}\rangle$ from the first 3-wave, the w-p $|1;\mathbf{b}\rangle$ from the second 3-wave, and the w-p $|1;\mathbf{c}\rangle$ from the third 3-wave. The w-p $|1;\mathbf{a}\rangle$ carries the phase-shift $\theta_1$, the w-p $|1;\mathbf{b}\rangle$ carries the phase-shift $\theta_2$, and the w-p $|1;\mathbf{c}\rangle$ carries the phase-shift $\theta_3$.

2) Choosing in (8) $\theta_1 = \theta_3 = 0$, and $\theta_2 = \pi$ there results

$$\text{Prob}(D_1 \& D_2 \& D_3) = |\mathcal{M}|^2 / 24. \tag{10}$$

Let's notice that introducing the transformation (6) in the 3-wave $|1;\mathbf{a}\rangle|1;\mathbf{e}_2\rangle|1;\mathbf{e}_3\rangle$, the contribution to $D_1 \& D_2 \& D_3$ is $|\mathcal{M}|^2 / 24$, the same as the result of all the three 3-waves together given by (10). If we

introduce (6) in the 3-wave $/1;\mathbf{e}_1\rangle/1;\mathbf{b}\rangle/1;\mathbf{e}_3\rangle$, the contribution to $D_1\&D_2\&D_3$ is also $|\mathcal{M}|^2/24$. If we introduce (6) in the 3-wave $/1;\mathbf{e}_1\rangle/1;\mathbf{e}_2\rangle/1;\mathbf{c}\rangle$, the contribution to $D_1\&D_2\&D_3$ is again $|\mathcal{M}|^2/24$.

This is the consequence of the three-particle interference: two of the amplitudes on the RHS of (8) cancel out mutually, and only one remains.

## 4. The assumption of the existence of a particle

Let's now see what would happen if some of the w-ps would carry particles, and other wave-packets were "empty waves".[2]

In a given trial of the experiment the S-particle would travel either with the w-p $/1;\mathbf{a}\rangle$, or with $/1;\mathbf{b}\rangle$, or with $/1;\mathbf{c}\rangle$. It is assumed below that the S-particle travels with a certain w-p, that it does not jump from one w-p to another, i.e. it follows a continuous trajectory. For instance, if the S-particle resides in the w-p $/1;\mathbf{c}\rangle$, there it remains all the way from $BS'$ to $BS_3$.

Option **a)** If the w-p $/1;\mathbf{a}\rangle$ carries the S-particle, and $/1;\mathbf{e}_2\rangle$ and $/1;\mathbf{e}_3\rangle$ carry, each one, a particle from the respective oscillator, then the 3-wave $/1;\mathbf{a}\rangle/1;\mathbf{e}_2\rangle/1;\mathbf{e}_3\rangle$ can produce a response $D_1\&D_2\&D_3$. The probability that $/1;\mathbf{a}\rangle$, $/1;\mathbf{e}_2\rangle$, and $/1;\mathbf{e}_3\rangle$, carry, each one, a particle, was found in the former section equal to $|\mathcal{M}|^2/3$, as one can get from (4). Therefore, the joint probability that the S-particle goes to $/1;\mathbf{a}\rangle$ and then to $D_1$, an oscillator particle goes to $/1;\mathbf{e}_2\rangle$ and then to $D_2$, and another oscillator particle goes to $/1;\mathbf{e}_3\rangle$ and then to $D_3$, is equal to

$$\text{Prob}_\mathbf{a}(D_1\&D_2\&D_3) = |\mathcal{M}|^2/24. \quad (11)$$

Option **b)** Analogously, if $/1;\mathbf{b}\rangle$ carries the S-particle, and $/1;\mathbf{e}_1\rangle$ and $/1;\mathbf{e}_3\rangle$ carry, each one, a particle from the respective oscillator, then the 3-wave $/1;\mathbf{e}_1\rangle/1;\mathbf{b}\rangle/1;\mathbf{e}_3\rangle$ could produce a response $D_1\&D_2\&D_3$, and the probability for that would be the same as that computed for the option **a**,

$$\text{Prob}_\mathbf{b}(D_1\&D_2\&D_3) = |\mathcal{M}|^2/24. \quad (12)$$

Option **c)** If $/1;\mathbf{c}\rangle$ carries the S-particle, and $/1;\mathbf{e}_2\rangle$ and $/1;\mathbf{e}_3\rangle$ carry, each one, a particle from the respective oscillator, then the 3-wave $/1;\mathbf{e}_1\rangle/1;\mathbf{e}_2\rangle/1;\mathbf{c}\rangle$ could produce a response $D_1\&D_2\&D_3$ analogously with the previous cases,

$$\text{Prob}_\mathbf{c}(D_1\&D_2\&D_3) = |\mathcal{M}|^2/24. \quad (12)$$

---

[2] The terminology "full" and "empty" waves is used in the literature for denoting a wave that impresses, respectively, does not impress, a detector. In Bohm's mechanics a full wave carries a particle that impresses a detector, while an empty wave is supposed to participate in interference but it carries no particle and thus, it does not impress a detector. A wide literature was written for testing theoretically and experimentally the existence of empty waves, see [22] – [28].

Since the result $D_1 \& D_2 \& D_3$ would be obtained either through option **a**, or through option **b**, or through **c**, the total probability of getting this result would be $|\mathcal{M}|^2/8$, which disagrees with the quantum prediction (10) by which the total probability of $D_1 \& D_2 \& D_3$ is $|\mathcal{M}|^2/24$.

## 5. Discussion and conclusions

A reader of this text argued that it does not exclude the concept of substructure particles. He said that it is just an example of wave-behavior, while other experiments exemplify the particle behavior.
I agree, it's not the concept of substructure particles that is ruled out here, but the possibility of continuous trajectories for them. If so, one can think that there exists substructure particle in discontinuous movement, as in the interpretation proposed by S. Gao. Though, as said in section 1, his proposal is problematic when examined vs. entanglements.

The impossibility of continuous trajectories for quantum particles cast a doubt over the dB-B interpretation of the QM, of which one of the basic principles is the continuity of such trajectories. If quantum objects have trajectories, the single change that these trajectories may undergo when passing from a frame of coordinates to another, should be the transformation according to the Lorentz transformation.
Though, some people are not satisfied with Hardy's thought-experiment since it involves moving frames. There are claims that if the relativity is invoked, the quantum fields theory should be used [29]. I disagree, since the experiment in [5] doesn't need high velocities. Though, for respecting the doubts in [29], the thought-experiment presented here works with a single frame of coordinates.

### Acknowledgement

I am in debt to prof. K. Kassner who insisted that this proof does not rule out the concept of particle. I agree with him, this proof rules out only the continuous trajectories for substructure particles.

---